\documentclass[journal=jpccck,manuscript=article]{achemso}

\usepackage{chemformula} 
\usepackage[T1]{fontenc} 
\usepackage{graphicx}

\usepackage{color}
\usepackage[bookmarksopen=true,bookmarksnumbered=true,colorlinks=true,urlcolor=blue,linkcolor=blue,citecolor=blue]{hyperref}

\usepackage{cleveref}

\author{Yu Chen}
\author{Shengxiang Wu}
\author{Shiwu Gao}
\affiliation{Beijing Computational Science Research Center, Beijing, 100193, China}
\email{swgao@csrc.ac.cn}

\title{Unified description of thermal and nonthermal hot carriers in plasmonic photocatalysis}

\begin{document}
		
	\begin{tocentry}
		\begin{center}
			\includegraphics{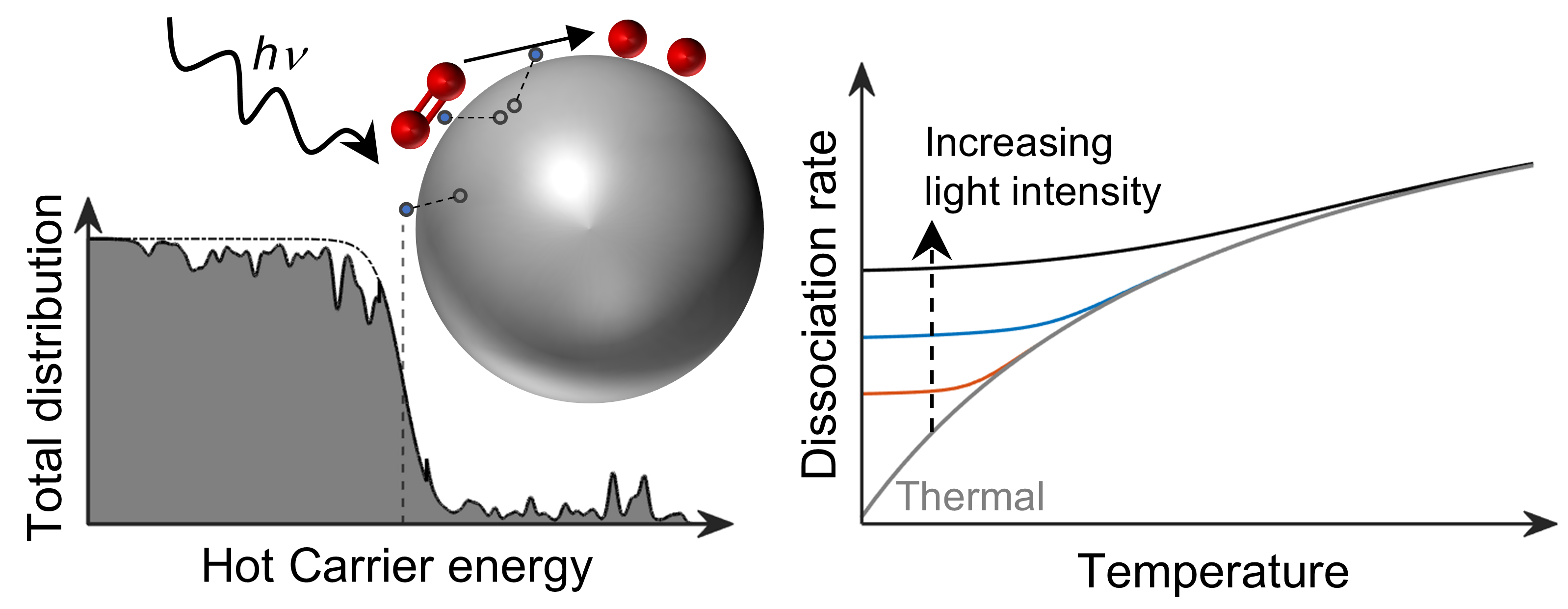}
		\end{center}
	\end{tocentry}
	
	\begin{abstract}
		The damping of surface plasmons generates hot carriers, which holds promise for photoelectric conversion and photocatalysis. Recent studies have revealed the nonequilibrium characters of the plasmonic hot carriers and their nonadiabatic coupling to molecular vibrations. Yet, the precise mechanism of plasmonic photocatalysis remains elusive and controversial. Here, we present a unified description of thermal and nonthermal hot carriers in the dynamics of vibrational excitation and photodissocation, where a quantitative comparison between the two mechanisms can be made. We revisit a well-studied system, O$_2$ dissociation on silver nanoparticles. The nonthermal hot carriers are found to promote molecular dissociation in the low-temperature or high-intensity regime. With increasing laser intensity, the dissociation rate exhibits a transition from a linear to nonlinear dependence due to the onset of vibrational heating as observed in experiment. Our model provides a unified framework to understand the mechanism and dynamics of photocatalysis and reveals the nonthermal pathways for energy
		harvesting and conversion with surface plasmons.
	\end{abstract}
	
	\section{Introduction}
	Since the pioneering works of plasmon-induced dissociation of O$_2$/H$_2$ on noble metal nanoparticles,~\cite{Christopher2011,Christopher2012,Mukherjee2013,Mukherjee2013_2} plasmonic photochemistry has evolved into an active research field with promising energy applications.~\cite{Moskovits2013,Linic2017,Maier2017,Xiong2019,Nordlander2016,Nordlander2019,Nordlander2020,Nordlander2018,Cortes2023,Cortes2024,Jain2018,Nordlander2020_2,Cortes2024_2,Linic2016_1,Kim2018,Nordlander2019_2,Dubi2020,Dubi2022,Wu2022,Meng2016,Meng2022,Hua2018,Hua2020,Linic2018review,Tian2020review} The hot carriers generated by Landau damping of surface plasmons have been suggested to be the dominant mechanism~\cite{Christopher2011,Christopher2012,Mukherjee2013,Mukherjee2013_2,Moskovits2013,Linic2017,Maier2017,Xiong2019,Nordlander2016,Nordlander2018,Nordlander2019,Nordlander2020,Cortes2023,Cortes2024,Jain2018,Nordlander2020_2,Cortes2024_2,Linic2016_1}, due to the efficient injection and separation of plasmonic hot carriers across the interface.~\cite{Tachiya2007,Lian2013,Lian2015,Atwater2020,Yang2024,Link2024,Gao2019,Gao2021,Erhart2019_1,Erhart2019_2} Such an electronic mechanism is common in electron- and photochemistry on solid surfaces, which is inevitably competed by possible thermal heating at ambient conditions and under irradiations.~\cite{Dubi2019,Dubi2020,Dubi2022,Quidant2020} This competition can be more complicated at the plasmonic hot spots of nanoparticle aggregates, where local heating~\cite{Govorov2006,Baffou2013} is hard to measure experimentally.~\cite{Dubi2020,Quidant2020} Recently, progresses have been made to disentangle the effects of hot carriers from photothermal heating using improved control measurements of surface temperatures,~\cite{Nordlander2021,Liu2018,Tian2019,Tian2021,Baldi2020,Miyar2025} yet the debate between thermal and nonthermal mechanisms remains inconclusive.
	
	Theoretical studies of plasmon-induced hot carrier generation based on electron gas models~\cite{Govorov2013,Manjavacas2014,Liu2017,Abajo2016,Govorov2017} or atomistic simulations~\cite{Govorov2015,Govorov2020,Louie2015,Narang2014,Narang2015,Atwater2017,Lischner2019,Lischner2022,Lischner2023,Erhart2020,Erhart2024} have shown highly nonthermal characters in the energy distributions. Whether and how these nonthermal hot carriers participate directly in surface photochemistry remains challenging for first-principles time-dependent density functional theory (TDDFT) simulations~\cite{Meng2016,Meng2022,Hua2018,Hua2020}. Phenomenological description of molecular dissociation has been based exclusively on Arrhenius laws with modified energy barriers and/or local temperatures~\cite{Nordlander2018,Nordlander2019,Nordlander2020,Dubi2020,Dubi2022}, which literally assume quasi-equilibrium distributions achieved in the electronic and/or vibrational degrees of freedom. Such an assumption contradicts with more recent experiments and TDDFT calculations, which show: 1) highly nonthermal characters of hot carrier injection and separation at the metal-semiconductor interface in the first-principles simulations,~\cite{Gao2019,Gao2021,Erhart2019_2} 2) ultrafast timescale of charge transfer in comparison with the much longer thermalization time within the nanoparticles,~\cite{Lian2015,Atwater2020,Yang2024,Link2024} 3) the nonthermal hot carrier distributions in direct optical measurements,~\cite{Heilpern2018,Wu2018,Edgar2020,Budai2022,Kumagai2023} 4) the nonthermal vibrational populations~\cite{Kim2023,Kim2024} and mode-specific vibrational temperatures of adsorbed molecules.~\cite{Linic2016_1,Linic2016_2,Frontiera2018,Dong2024} All these observations demonstrate the nonequilibrium nature of the plasmonic hot carriers and the induced vibrational excitations, which remain to be understood. 
	
	The Anderson-Newns Hamiltonian~\cite{Anderson1961,Newns1969} with a linear electron-vibration coupling to truncated harmonic oscillators has been appealing to effectively account for the dynamical activation of molecular vibrations by plasmonic hot carriers,~\cite{Nordlander2024,Christopher2014,Alabastri2024,Kim2023,Kim2024} treating the bond breaking in the DIET (desorption induced by electronic transition~\cite{DR1964,Redhead1964}) and/or DIMET (desorption induced by multiple electronic transitions~\cite{Newns1992}) mechanisms. For example, Linic and coworkers have adopted this model to explain their experimental findings~\cite{Christopher2011,Christopher2012} including the light intensity dependence and the kinetic isotope effect. Unfortunately, the invoked formulation~\cite{Olsen2009} considered only a single inelastic scattering event with fixed carrier energy at the molecular resonance, and literally omitted all possible contributions from hot carriers in a wide energy spectrum with nonthermal distributions, which have been identified by numerous TDDFT simulations ~\cite{Gao2019,Gao2021,Erhart2020,Erhart2024} and the latest experimental measurements.~\cite{Heilpern2018,Wu2018,Edgar2020,Budai2022,Kumagai2023}. As a result, the effect of nonthermal hot carriers on photocatalysis, and its possible competition with thermal carriers and photothermal heating mechanism~\cite{Nordlander2018,Dubi2020,Dubi2022,Quidant2020,Nordlander2021,Liu2018,Tian2019,Tian2021,Baldi2020,Miyar2025} remain unresolved.
	
	We have recently developed a unified description of vibrational excitation induced by both thermal and nonthermal hot carriers,~\cite{Wu2022} where an explicit form of nonthermal distribution of plasmonic hot carriers has been introduced. It allows a quantitative comparison between nonthermal and thermal excitation mechanisms. This paper presents a comprehensive description of the model and important extensions to separate the electron and hole channels of nonthermal excitations, and to incorporate the vibrational heating in a wide temperature and light intensity regimes. The mode-specific electron-vibration coupling provides direct physical insight into the mode-dependent vibrational excitations observed in recent experiments.~\cite{Linic2016_1,Linic2016_2,Frontiera2018,Dong2024,Kim2023,Kim2024} As an application of this model, we reanalyzed O$_2$ dissociation on silver nanoparticle (AgNP) as functions of light intensities and temperatures. We found that beyond a moderate light intensity, local vibrational heating through multiple electronic scatterings outweighed the DIET mechanism for plasmonic photocatalysis. It reproduced the nonlinear dissociation rates at higher light intensities.~\cite{Christopher2012,Mukherjee2013,Nordlander2016,Nordlander2019,Nordlander2020} Our work provides a unified framework to understand hot carrier photochemistry starting from the optical absorption and electron-vibration coupling, and bridges the gap between the thermal and nonthermal mechanisms for plasmonic photocatalysis.
	
	\section{Methods}
	\subsection{Hamiltonian}
	The nanoparticle-molecule system is described by the Anderson-Newns model~\cite{Anderson1961,Newns1969} with a linear electron-vibration coupling to a truncated harmonic oscillator~\cite{PP1980,Gadzuk1991,Gao1997,SurSci1,SurSci2,Gao1997tunnel,Olsen2009}. The total Hamiltonian is comprised of the following three parts: 
	\begin{equation}\label{eqn1}
		H=H_e+H_v+H_i,
	\end{equation}
	where
	\begin{subequations}
		\begin{align}
			H_e&=\varepsilon_a c_a^\dagger c_a+\sum_k \varepsilon_k c_k^\dagger c_k+\sum_k (v_{ak} c_a^\dagger c_k+\mathrm{H.c.}),\label{eqn2a}\\
			H_v&=\hbar\Omega b^\dagger b,\label{eqn2b}\\
			H_i&=\lambda c_a^\dagger c_a(b^\dagger+b)\label{eqn2c}.
		\end{align}
	\end{subequations}
	The electronic Hamiltonian $H_e$ consists of a molecular resonance $|a\rangle$ with energy $\varepsilon_a$, the continuum states of the metal nanoparticles $|k\rangle$ with energies $\varepsilon_k$, and the tunneling coupling $v_{ak}$ between them. The molecular vibration $H_v$ is approximated by a harmonic oscillator with vibrational frequency $\Omega$.
	
	In the resonance electron-vibration coupling, vibrational excitation is induced by the temporary occupation of the resonance state ($n_a=c_a^\dagger c_a$), resulting in vibrational motion on an excited state, which is described by a shifted harmonic oscillator with reaction coordinate $q$. The coupling constant $\lambda$ for the linear electron-vibration interaction is determined by
	\begin{equation}\label{eqn3}
		\lambda=\sqrt{\frac{\hbar}{2\mu\Omega}}\frac{d}{dq}\varepsilon_a(q)\Big|_{q=q_0},
	\end{equation}      
	where $\mu$ is the reduced mass for the oscillator, and $q_0$ corresponds to the equilibrium position of the ground state. This term is mode-dependent, which may lead to mode-specific vibrational excitations in adsorbed molecules as demonstrated in early calculations~\cite{Gao1994} and recent experiments.~\cite{Linic2016_1,Linic2016_2,Frontiera2018,Dong2024,Kim2023,Kim2024} Such a model Hamiltonian has been generally used to describe various surface dynamical processes in surface sciences, including vibrational damping~\cite{PP1980}, electron energy loss spectroscopy~\cite{Gadzuk1991}, dissociation induced by tunneling electrons~\cite{Ho1997,Gao1997tunnel}, and femtosecond laser induced desorption and dissociations~\cite{Gao1997,SurSci1,SurSci2,Olsen2009}.
	
	\subsection{Vibrational transition rates induced by plasmonic hot carriers}
	To describe plasmon induced photocatalysis under CW laser irradiation, it is essential to consider the nonthermal hot carriers that are constantly generated during plasmonic damping. These hot carriers can couple to the molecule on a timescale given by the tunneling lifetime of the resonance state, which is typically a few femtoseconds~\cite{Lian2015,Atwater2020,Gao2019,Gao2021,Yang2024,Link2024}. It is much shorter than the thermalization time on the picoseconds for hot carriers in metals and within the nanoparticles~\cite{ee1999,ee2000,eph1995,eph2003,Atwater2017}. The persistent generation of hot carriers by CW lasers makes it possible to maintain a nonthermal distribution on top of the thermal distribution as governed by the Fermi-Dirac distribution $f_{\mathrm{0}}(\varepsilon, T)$ with an electron temperature. The nonthermal distribution function under CW illumination is defined as~\cite{Wu2022}
	\begin{equation}\label{eqn4}
		f_{\mathrm{nth}}(\varepsilon)=\frac{\delta\rho(\varepsilon)}{\rho_{\mathrm{DOS}}(\varepsilon)}.
	\end{equation} 
	Here $\delta\rho(\varepsilon)$ is the density response of hot carriers generated by the Landau damping of plasmons following Govorov~\cite{Govorov2013,Govorov2015} as
	\begin{equation}\label{eqn5}
		\begin{split}
			\delta\rho(\varepsilon)&=4e^2\sum_{kk^\prime}\frac{f_0(\varepsilon_{k^\prime})-f_0(\varepsilon_k)}{\Gamma_{kk}}|V_{kk^\prime}|^2\left[\frac{\Gamma_{kk^\prime}}{(\hbar\omega-\varepsilon_k+\varepsilon_{k^\prime})^2+\Gamma_{kk^\prime}^2}\right.\\&\left.\hspace{5em}+\frac{\Gamma_{kk^\prime}}{(\hbar\omega+\varepsilon_k-\varepsilon_{k^\prime})^2+\Gamma_{kk^\prime}^2}\right]\delta(\varepsilon-\varepsilon_k),
		\end{split}
	\end{equation}
	with $V_{kk^\prime}$ the optical transition matrix element, $\hbar\omega$ the energy of incident photon, and $\Gamma_{kk^\prime}$=$\Gamma_{kk}$=$\hbar/\tau$ the relaxation matrix element in the relaxation time approximation. This $f_{\mathrm{nth}}$ can be treated on equal footing with the equilibrium counterpart $f_0$. 
	
	For an electron bath with a general distribution function $f(\varepsilon)$, the inelastic transition rate from vibrational state $|n\rangle$  to state $|n^\prime\rangle$ of the admolecule can be evaluated as~\cite{Gadzuk1991,Gao1997,Gao1997tunnel}
	\begin{equation}\label{eqn6}
		\begin{split}
			W_{n\to n^\prime}&=\frac{4\Delta_a^2}{\pi\hbar}\int\mathrm{d}\varepsilon f(\varepsilon)[1-f(\varepsilon+\delta\varepsilon)]|T_{nn^\prime}(\varepsilon)|^2\\
			&=\frac{4\Delta_a^2}{\pi\hbar}\int\mathrm{d}\varepsilon f(\varepsilon)[1-f(\varepsilon+\delta\varepsilon)]\bigg|\sum_m \frac{\langle n^\prime|m\rangle\langle m|n\rangle}{\varepsilon+\varepsilon_n-\tilde{\varepsilon}_a-\varepsilon_m+i\Delta_a}\bigg|^2,
		\end{split}
	\end{equation}
	where $T_{nn^\prime}$ is the transition amplitude between two bound levels $|n\rangle$ and $|n^\prime\rangle$ via intermediate states $|m\rangle$, $\varepsilon_n=n\hbar\Omega$ is the vibrational energy of state $n$, $\delta\varepsilon=\varepsilon_n-\varepsilon_{n^\prime}$ is the energy transfer between electrons and the molecule. $\Delta_a=\sum_k \pi v_{ak}^2\delta(\varepsilon-\varepsilon_k)$ is the broadening of molecular resonance, which effectively accounts for the tunneling coupling to the electron bath. Its Hilbert transform $\Lambda_a=\sum_k \int \mathrm{d}\varepsilon\frac{v_{ak}^2}{\varepsilon-\varepsilon_k}$ yields the energy shift of the resonance state $\tilde{\varepsilon}_a=\varepsilon_a+\Lambda_a$. 
	
	For an electron bath with a thermal Fermi-Dirac distribution at electron temperature $T_e$, it is not necessarily in equilibrium with the lattice or environment in a dynamical process. The integration over the energy distribution can be evaluated in the wide-band approximation~\cite{Gadzuk1991,Gao1997,Gao1997tunnel}, using the analytic expression of the Franck-Condon factors between two displaced harmonic oscillators~\cite{Iachello1998}. The result reads
	\begin{equation}\label{eqn7}
		\begin{split}
			W_{n\to n^\prime}^{\mathrm{th}}=&\frac{4\Delta_a^2(n^\prime-n)\Omega}{\pi}n_\mathrm{B}(T_e,(n^\prime-n)\hbar\Omega)\\&\times\bigg| \sum_{m}\dfrac{F(m,n,n^\prime;\beta)}{\varepsilon_\mathrm{F}+(n^\prime+n-2m)\hbar\Omega/2-\tilde{\varepsilon}_a+i\Delta_a}\bigg|^2,
		\end{split}
	\end{equation}
	where $n_\mathrm{B}(T_e,\varepsilon)=[\mathrm{exp}(\varepsilon/k_\mathrm{B} T_e)-1]^{-1}$ is the Bose-Einstein distribution at $T_e$, $\beta=\lambda/(\hbar\Omega)$, and 
	\begin{equation}\label{eqn8}
		\begin{split}
			F(m,n,n^\prime;\beta)=&\frac{\sqrt{n!n^\prime!}}{m!}e^{-\beta^2}\beta^{n+n^\prime+2m}\\
			&\times\sum_{k_1=0}^{\mathrm{min}[n^\prime,m]}\sum_{k_2=0}^{\mathrm{min}[n,m]}\frac{\binom{m}{k_1}\binom{m}{k_2}}{(-\beta^2)^{k_1+k_2}(n^\prime-k_1)!(n-k_2)!}.
		\end{split}
	\end{equation}
	Equation~\eqref{eqn7} guarantees the detailed balance between vibrational transitions, $W_{n\to n^\prime}^\mathrm{th}/W_{n^\prime\to n}^\mathrm{th}=\exp[(\varepsilon_{n}-\varepsilon_{n^\prime})/k_\mathrm{B}T_e]$. When the lattice and the electron bath are in equilibrium at the same temperature, equation~\eqref{eqn7} represents the vibrational transition rate without laser irradiation, namely under the dark condition.
	
	To incorporate the nonthermal hot carriers generated by plasmon damping, the electron distribution function can be generalized to include the nonthermal term as~\cite{Wu2022}
	\begin{equation}\label{eqn9}
		f(\varepsilon,T)=f_\mathrm{0}(\varepsilon, T)+f_{\mathrm{nth}}(\varepsilon),
	\end{equation}
	since these nonthermal hot carriers also participate in the molecular scattering and vibrational excitation. In order to separate the thermal and nonthermal contributions, the product between distribution functions in Eq.~\eqref{eqn6} is partitioned into four parts, 
	\begin{equation}\label{eqn10}
		\begin{split}
			f(\varepsilon)(1-f(\varepsilon+\delta\varepsilon))=&(f_\mathrm{0}(\varepsilon)+f_\mathrm{nth}(\varepsilon))[1-(f_\mathrm{0}(\varepsilon+\delta\varepsilon)+f_\mathrm{nth}(\varepsilon+\delta\varepsilon))]\\=&f_\mathrm{0}(\varepsilon)(1-f_\mathrm{0}(\varepsilon+\delta\varepsilon))+f_\mathrm{nth}(\varepsilon)(1-f_\mathrm{0}(\varepsilon+\delta\varepsilon))\\&-f_\mathrm{0}(\varepsilon)f_\mathrm{nth}(\varepsilon+\delta\varepsilon)-f_\mathrm{nth}f_\mathrm{nth}(\varepsilon+\delta\varepsilon).
		\end{split}
	\end{equation}
	Here the first term corresponds to the contribution from the thermal distributions. Under the normal laser intensity, $f_\mathrm{nth}$ should be much smaller compared to $f_\mathrm{0}$. Therefore the last term is much smaller and can be neglected. The second and third terms correspond to the contributions from the nonthermal hot electrons and hot holes, respectively. The vibrational transition rates can be reorganized as
	\begin{equation}\label{eqn11}
		\begin{split}
			W_{n\to n^\prime}^\mathrm{nth}&=W_{n\to n^\prime}^\mathrm{HE}+W_{n\to n^\prime}^\mathrm{HH},\\
			W_{n\to n^\prime}^\mathrm{HE}&=\frac{4\Delta_a^2}{\pi\hbar}\int\mathrm{d}\varepsilon\;f_\mathrm{nth}(\varepsilon)(1-f_0(\varepsilon+\delta\varepsilon))|T_{nn^\prime}(\varepsilon)|^2,\\
			W_{n\to n^\prime}^\mathrm{HH}&=-\frac{4\Delta_a^2}{\pi\hbar}\int\mathrm{d}\varepsilon\;f_0(\varepsilon)f_\mathrm{nth}(\varepsilon+\delta\varepsilon)|T_{nn^\prime}(\varepsilon)|^2.
		\end{split}
	\end{equation}
	Equation~\eqref{eqn11} can be numerically evaluated. It provides an unambiguous and concise way to calculate the nonthermal transition rates from hot electrons and holes, which was done by qualitatively dividing the distributions into three energy regimes~\cite{Wu2022}.
	
	\subsection{Reaction rates in DIET and DIMET regimes}
	In the truncated harmonic oscillator approximation, molecular dissociation occurs when the vibrational energy surpasses the dissociation barrier. By considering the vibrational transitions from all bound states up to the dissociation level, $n_\mathrm{d}$, the dissociation rate under laser excitation is given by:
	\begin{equation}\label{eqn12}
		R_{\mathrm{dis}}=\sum_{n=0}^{n_\mathrm{d}-1} p_n(T) W_{n\to n_{\mathrm{d}}},
	\end{equation}
	where $p_n(T)$ is the thermal population of vibrational state $n$ at temperature $T$. For reaction induced by single electronic transition (DIET regime), their vibrational populations follow exactly the Bose-Einstein statistics at electronic temperature $T_e$, which is assumed to be in thermal equilibrium with the environment. 
	
	The rate of dissociation in the DIET mechanism accounts only for a single inelastic scattering from a bound state. It does not include possible modification of vibrational populations by the multiple scatterings of nonthermal carriers in the DIMET regime. Experimentally, vibrational heating could occur under high laser irradiations~\cite{Hocrossover1997} or in plasmonic gaps with strong local field enhancement~\cite{Christopher2012,Mukherjee2013,Nordlander2016,Nordlander2019,Nordlander2020}. It can be described efficiently by introducing a modified vibrational temperature $T_\mathrm{v}$, which in principle should be determined by solving the Pauli master equation~\cite{Gao1997,Gao1997tunnel}. 
	
	A simple and approximate way to determine the vibrational temperature in DIMET regime can be obtained by extracting the temperature from the vibrational transitions between the ground state $n=0$ and the $n=1$ excited state, $W_{0\to 1}^\mathrm{tot}=\Gamma_\uparrow^\mathrm{tot}$ and $W_{1\to 0}^\mathrm{tot}=\Gamma_\downarrow^\mathrm{tot}$, 
	\begin{equation}\label{eqn13}
		T_\mathrm{v}=\dfrac{\hbar\Omega}{k_\mathrm{B}\ln\left(\dfrac{\Gamma_\downarrow^\mathrm{tot}}{\Gamma_\uparrow^\mathrm{tot}}\right)}=\dfrac{\hbar\Omega}{k_\mathrm{B}\ln\left(\dfrac{\Gamma_\downarrow^\mathrm{th}+\Gamma_\downarrow^\mathrm{nth}}{\Gamma_\uparrow^\mathrm{th}+\Gamma_\uparrow^\mathrm{nth}}\right)}.
	\end{equation}
	With the vibrational temperature, the dissociation rate within the DIMET mechanism can now be evaluated using the modified vibrational populations $p_n(T_\mathrm{v})$. We will later discuss the effect of the DIMET mechanism and make comparison with the DIET mechanism~\cite{Wu2022}, and compare with other theoretical descriptions available in the literatures~\cite{Christopher2012,Olsen2009}.
	
	\section{Results and Discussions \label{RD}}
	Here we apply the above model to study the vibrational excitation and dissociation of O$_2$ on AgNP, which has been intensively investigated experimentally~\cite{Christopher2011,Christopher2012}. Upon adsorption, the $\operatorname{O-O}$ bond lies parallelly to the surface \cite{Rocca1997,Schneider2000,Ho2005}, suggesting no direct coupling between molecular dipole and the local field. Hence, direct photolysis by electric field can be ruled out. To describe the electronic structures of AgNP, we adopt a jellium sphere model~\cite{Govorov2013,Manjavacas2014,Liu2017} for the ground state spectrum of spherical nanoparticles with a diameter up to $D= 25\;\mathrm{nm}$ and the Wigner-Seitz radius $r_s/r_\mathrm{B}=3.02$, where $r_\mathrm{B}$ is the Bohr radius. In the low coverage limit, the effect of molecular adsorption on the plasmon properties, the so-called chemical interface damping~\cite{Persson1993,Link2017,Link2019,Christopher2019,Link2024}, is minor and has been neglected. The adsorbate-specific parameters are extracted from first-principle calculations of O$_2$ chemisorbed on the Ag(100) \cite{Christopher2012}, with $\Delta_a=0.6\;\mathrm{eV}$, $\varepsilon_a=2.4\;\mathrm{eV}$ corresponding to the $2\pi^*$ resonance of the O$_2$ on Ag(100) surface formed by hot electron transfer, and $\hbar\Omega=0.10\;\mathrm{eV}$ being the vibrational energy of the adsorbed $\operatorname{O-O}$ stretch mode.
	
	\subsection{Hot carrier distribution \label{fnth}}
\begin{figure*}
		\includegraphics[width=3.3 in, keepaspectratio]{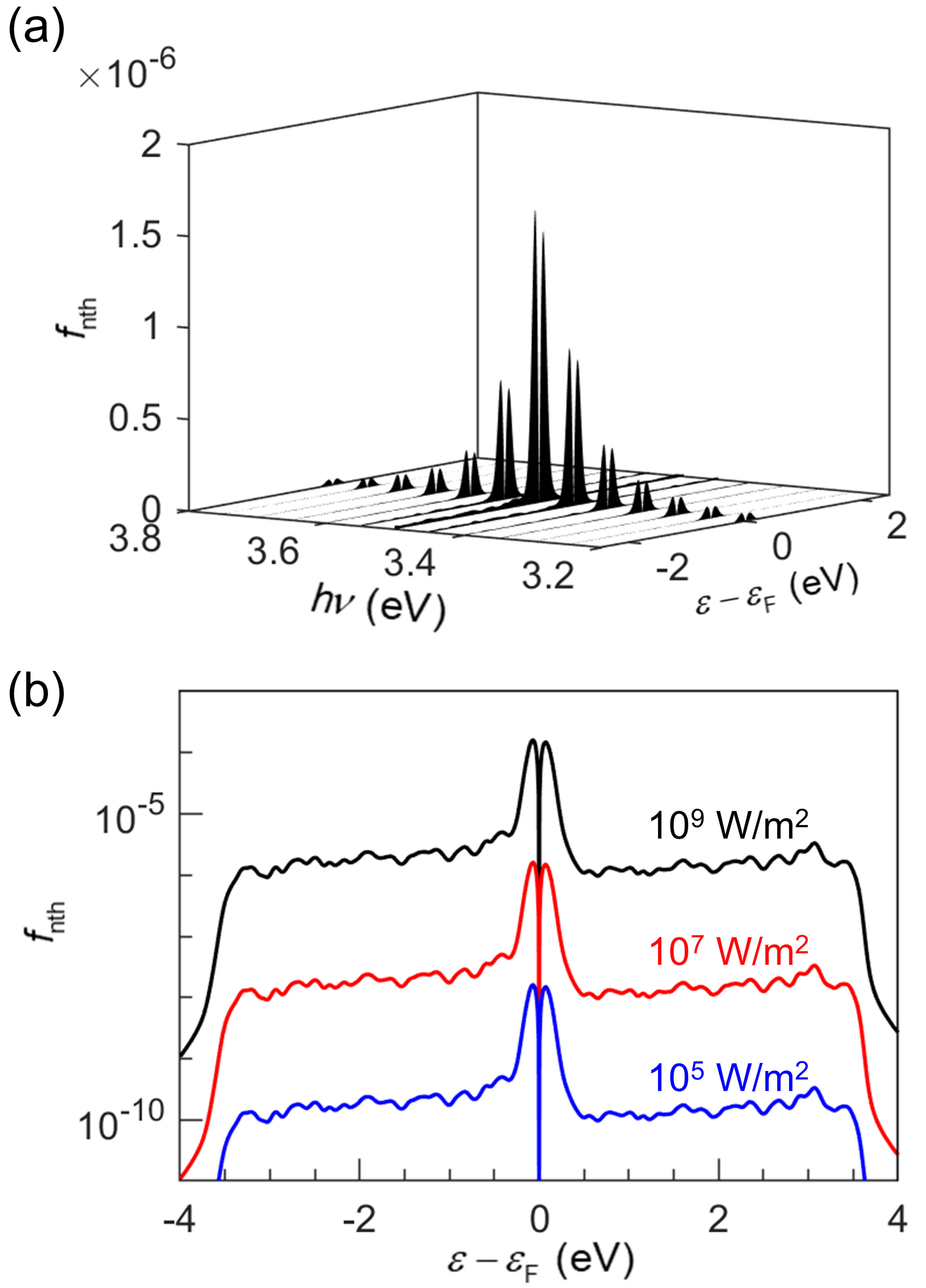}
		\caption{\label{fig:fig1} (a) Wavelength  and (b) light intensity dependences of the nonthermal energy distribution of plasmonic hot carriers. A significant resonance enhancement appears at $h\nu = 3.5\;\mathrm{eV}$, corresponding to the plasmon resonance of a silver sphere with diameter $D=25\;\mathrm{nm}$. Based on the linear response framework, the hot carrier generation depends linearly on light intensity. The calculations are taken at $T=300\;\mathrm{K}$, and $\tau = 100\;\mathrm{fs}$.} 
	\end{figure*}

	As a first step, the nonthermal distribution defined in Eq.~\eqref{eqn4} is first evaluated as a function of wavelength and laser intensity. Figure~\crefformat{figure}{#2#1{(a)}#3}\cref{fig:fig1} shows the energy and wavelength dependences of the nonthermal distribution function. When illuminated by light with given energy $h\nu$, the hot carriers generated are essentially distributed within the energy range $(\varepsilon_\mathrm{F},\varepsilon_\mathrm{F}+h\nu)$ for the electrons and $(\varepsilon_\mathrm{F}-h\nu,\varepsilon_\mathrm{F})$ for the holes, respectively. Both distributions of electrons and holes are plotted by the absolute magnitude. For a given particle size $D$ and carrier lifetime $\tau$, the nonthermal carriers are distributed predominantly near the Fermi level due to the available thermal electron-hole pairs near the Fermi surface. The hot carriers are therefore affected by the populations of thermal electron-hole pairs during Landau damping. Apart from this thermal character, there are purely nonthermal distributions at energies far away from the Fermi level. Increasing the carrier lifetime or reducing the nanoparticle size favors the hot carrier distribution and thus enhances the nonthermal distributions~\cite{Manjavacas2014}. In the dependence of photon energy, a clear resonance enhancement can be observed at $h\nu\sim3.5\;\mathrm{eV}$, which coincides with the calculated plasmon resonance $\hbar\omega_\mathrm{p}$ of the AgNP at the given diameter~\cite{Christy1972}. This resonance effect confirms that the plasmon excitation is dominant in hot carrier generations.
	
	Figure~\crefformat{figure}{#2#1{(b)}#3}\cref{fig:fig1} shows the hot carrier distributions as a function of the light intensity. These distributions have the same energy profiles. The magnitude of the distributions depends linearly on the light intensity $I_0$, as expected by the linear response theory~\cite{Dubi2019}. Under near-resonant excitation, the local field experienced by the conduction electrons is dominated by the plasmonic field, whose intensity scales linearly with that of the incident light~\cite{Govorov2013,Manjavacas2014,Liu2017}.
	
\begin{figure*}
		\includegraphics[width=3.3 in, keepaspectratio]{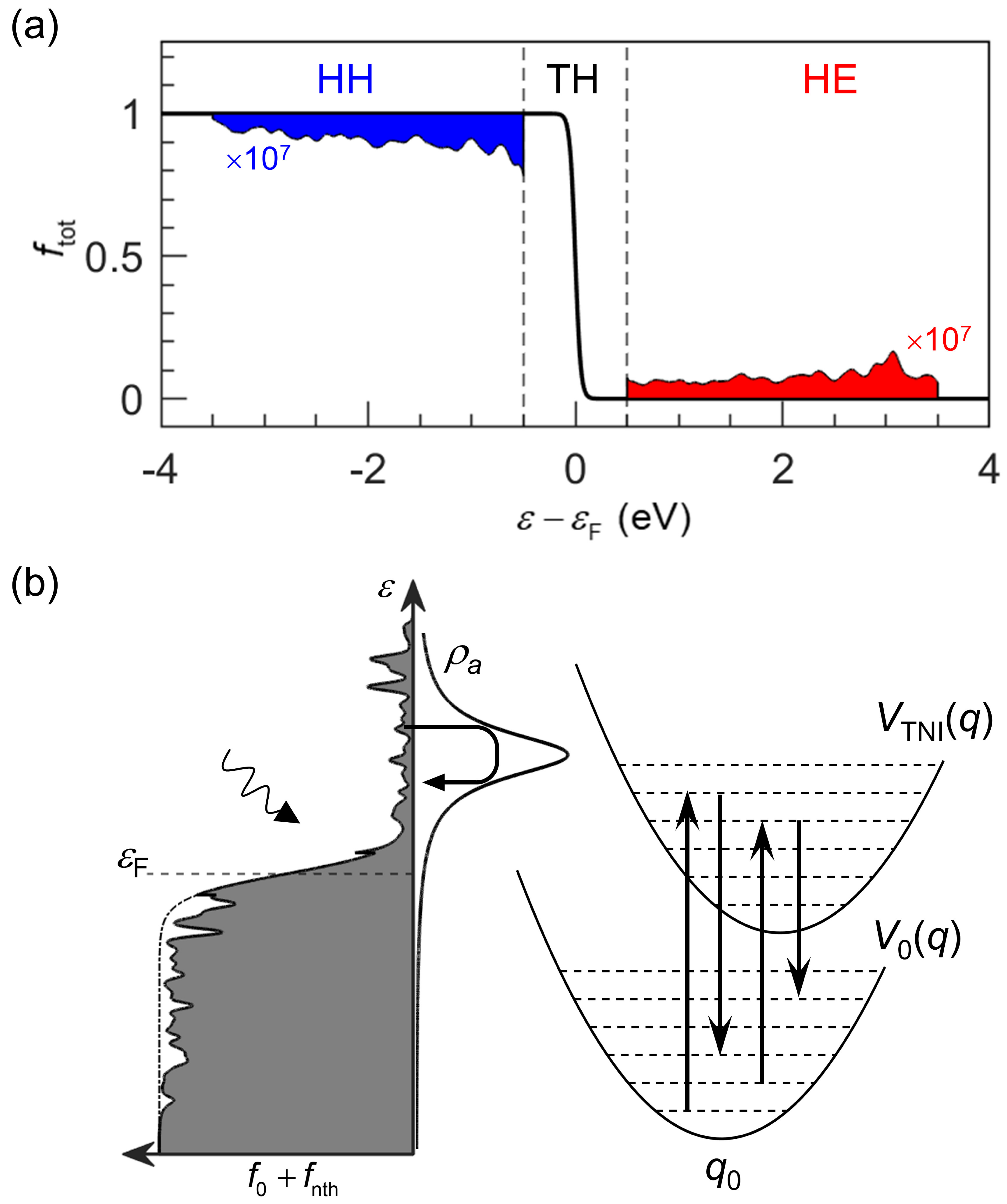}
		\caption{\label{fig:fig2} (a) The total electronic energy distribution at $T=300\;\mathrm{K}$ and $I_0=10^7\;\mathrm{W/m^2}$. Carriers are segmented into three regions. The TH region around the Fermi level contains mostly the thermal carriers, while the HH and HE regions consist of hot holes and hot electrons, respectively. The nonthermal characters are colored for clarification. (b) Schematic of a unified treatment of thermal and nonthermal hot carrier distributions in the vibrational heating mechanism.}
	\end{figure*}

	Figure~\crefformat{figure}{#2#1{(a)}#3}\cref{fig:fig2} shows in more detail the total hot carrier distribution function at $T=300\;\mathrm{K}$ including both the thermal distribution and the nonthermal distribution. The overall distribution is dominated by the thermal distribution, as expected. Nonthermal distributions are substantially smaller in magnitude for the low-energy electrons and holes near the Fermi level. However, away from the Fermi level, the thermal distributions exponentially decay while the nonthermal distributions do not. It implies that the high-energy distributions should always be dominated by the nonthermal electron-hole pairs. To help understand the different contributions from the thermal and nonthermal carriers, we could qualitatively divide the total distribution function into three regions. The thermal (TH) region near the Fermi level contains both thermal and nonthermal components, but is overwhelmingly dominated by the thermal carriers. The hot hole (HH) and hot electron (HE) regions predominantly consist of hot holes and hot electrons, respectively. The insets are close ups of distributions in the HH and HE regions.
	
	\subsection{Rates of vibrational transitions \label{wmn}}
\begin{figure*}
		\includegraphics[width=3.3 in, keepaspectratio]{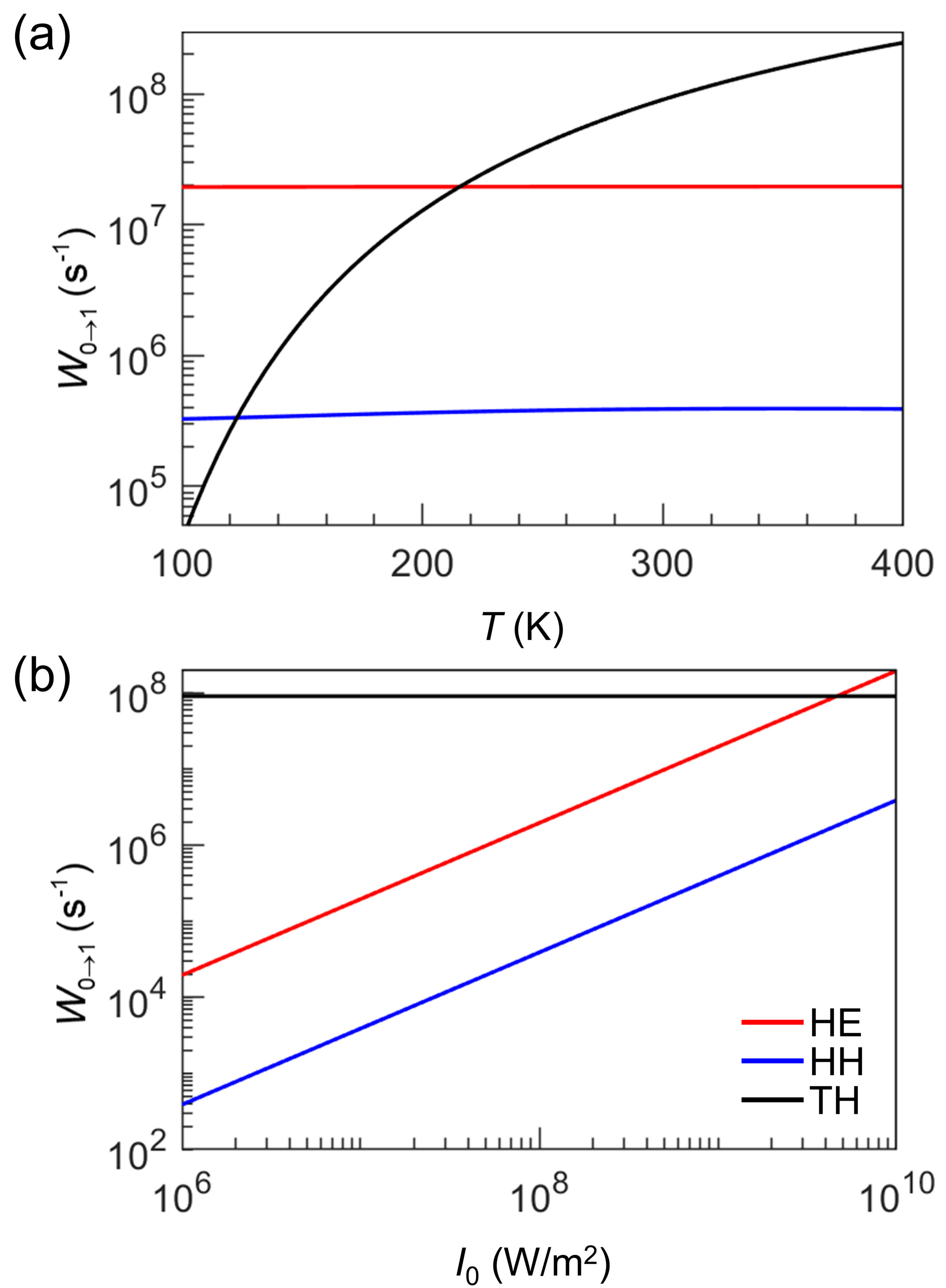}
		\caption{\label{fig:fig3} (a) Temperature and (b) light intensity dependences of the vibrational excitation rate $W_{0\to1}$ with contributions from hot electrons (red lines), hot holes (blue lines), and thermal carriers (black lines). The nonthermal carriers are efficient in promoting the activation of molecular vibrations in the low temperature and/or the high light intensity regime.}
	\end{figure*}

	To explore how the nonthermal hot carriers affect the vibrational excitation of the molecules adsorbed on the nanoparticle surface, whose general picture involving the tunneling coupling is schematically shown in~\crefformat{figure}{Fig.~#2#1{(b)}#3}\cref{fig:fig2}, we plot in~\crefformat{figure}{Fig.~#2#1{(a)}#3}\cref{fig:fig3} the rate of vibrational excitation $W_{0\rightarrow 1}$ as a function of temperature with separate contributions from hot electrons (red line), hot holes (blue line) and thermal carriers (black line). The thermal contribution increases with temperatures, while those of hot electrons and hot holes are temperature independent due to their nonthermal characters. The excitation by nonthermal hot carriers dominates at low temperatures, and is outweighed by thermal contribution at temperatures above 220~K. Moreover, the unoccupied O$_2$ $2\pi^*$ resonance favors obviously the vibrational coupling with hot electrons than the hot holes, leading to larger vibrational excitation rates for the hot electrons.
	
	Figure~\crefformat{figure}{#2#1{(b)}#3}\cref{fig:fig3} shows the light intensity dependence of $W_\mathrm{0\to1}$ at a fixed temperature of 300~K. Contributions from different carriers are presented in the same way as shown in Fig.~\crefformat{figure}{#2#1{(a)}#3}\cref{fig:fig3}. Here, the nonthermal contributions increase linearly with the light intensity as expected from the linear response theory. It is clear that the thermal excitation dominates $W_{0\to1}$ under weak illumination, and is overtaken by hot electrons for light intensities higher than~$\sim$4$\times$10$^9$~W/m$^2$. For vibrational transitions involving larger energy transfers, the nonthermal hot carriers can dominate vibrational excitations at much lower light intensities.
	
	\subsection{Dissociation rate and vibrational heating \label{rdis}}
\begin{figure*}
		\centering
		\includegraphics[width=3.3 in, keepaspectratio]{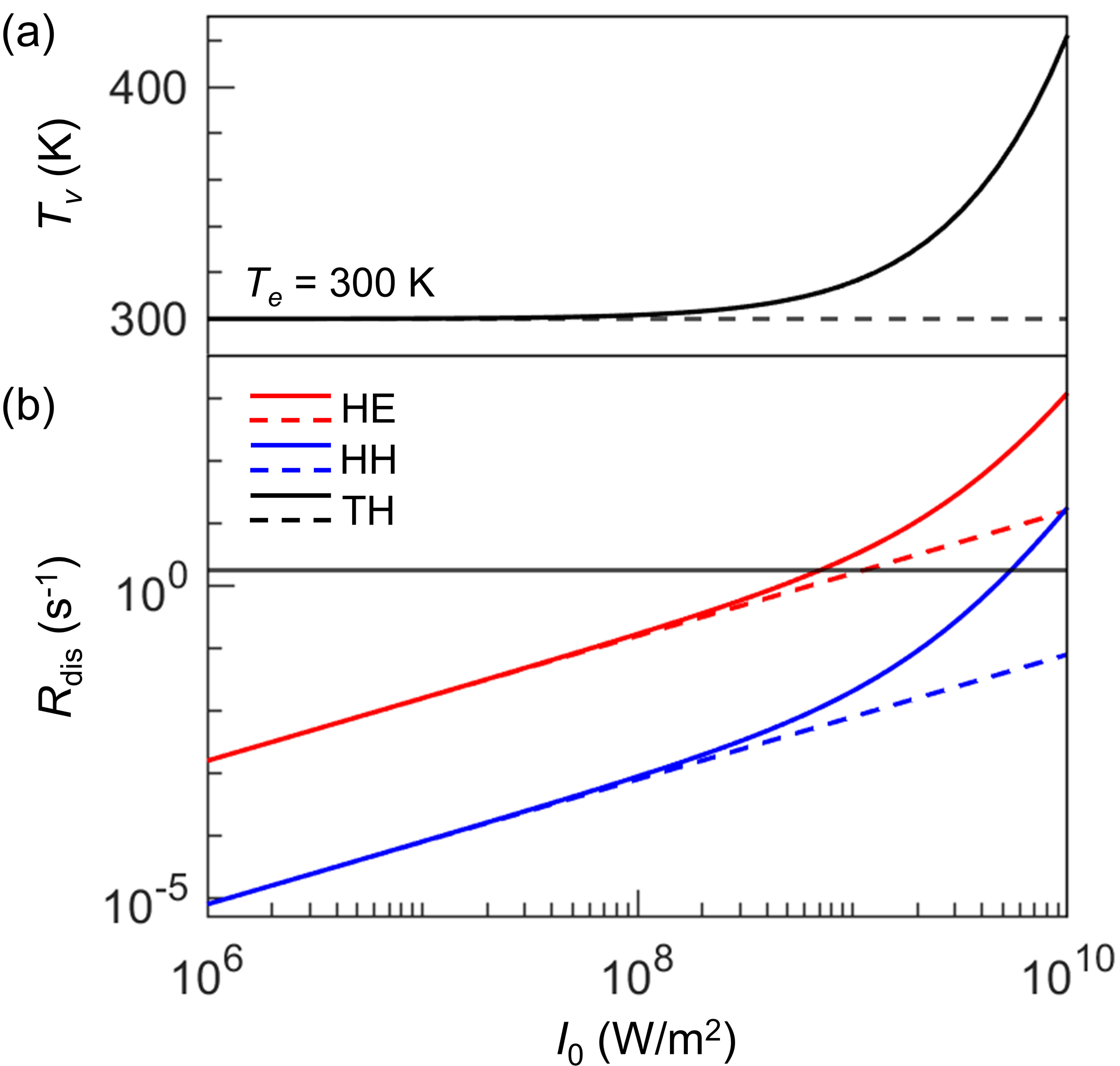}
		\caption{\label{fig:fig4} (a) Light intensity dependence of the vibrational temperature (solid lines) under resonant illumination. As compared to the electronic temperature (dashed lines), dramatic vibrational heating of admolecule can be observed especially at high values of light intensity. (b) The dissociation rates as a function of light intensity in the DIMET regime (solid lines) and are compared to the corresponding DIET curves (dashed lines). The transition from a linear to superlinear dependence can be observed, indicating the onset of vibrational heating.}
	\end{figure*}

	We next analyze the dissociation rate of O$_2$ on AgNP. A laser with $h\nu$~=3.5~{eV} is used for resonance plasmon excitation and the dissociation level is taken as $n_\mathrm{d}$~= 6. In our previous work~\cite{Wu2022}, only the DIET regime with a fixed light intensity has been considered. Yet, vibrational heating has been reported at high light intensity in the same system~\cite{Christopher2012}. To account for the modified vibrational distribution induced by multiple electronic scatterings, a vibrational temperature approximation~\cite{SurSci1,Gao1997} can be introduced. In~\crefformat{figure}{Fig.~#2#1{(a)}#3}\cref{fig:fig4}, the vibrational temperatures $T_\mathrm{v}$ given by Eq.~\eqref{eqn13} are displayed as a function of light intensity at fixed environmental temperature $T_\mathrm{env}=T_e=$~300~K. As shown in~\crefformat{figure}{Fig.~#2#1{(b)}#3}\cref{fig:fig3}, the vibrational excitations are still dominated by the thermal activation at low light intensities. The vibrational temperatures are thus in equilibrium with the electronic temperature. However, as the light intensity increases, the populations on higher vibrational states are increased consequently, which results in elevated vibrational temperatures. Indeed, mode-specific vibrational temperatures in adsorbed molecules has been reported using Raman thermometry~\cite{Linic2016_1,Linic2016_2,Frontiera2018,Dong2024,Kim2023,Kim2024}, and has been suggested as a signature of the hot carrier mechanism in plasmonic photocatalysis.
	
	The modified vibrational temperature leads to enhanced rates of dissociation especially in the DIMET regime. Figure~\crefformat{figure}{#2#1{(b)}#3}\cref{fig:fig4} shows the light intensity dependence of the dissociation rates with separate contributions from thermal and nonthermal hot carriers. Both the DIET (dashed lines) and the DIMET (sold lines) mechanisms are shown. For the thermal channel (TH), the vibrational temperature remains unperturbed ($T_\mathrm{v}=T_e$), resulting in identical rates in both mechanisms. The thermal electronic distribution is independent of the light intensity, leading to a constant dissociation rate. In contrast, nonthermal hot carriers lead to distinct intensity dependence of the dissociation rates. At low intensities, dissociation is governed by the DIET mechanism with a linear intensity dependence. As the light intensity increases, a transition from the linear to superlinear dependence occurs due to the onset of the vibrational heating. The superlinearity results from the additional vibrational transitions from higher vibrational states and the elevated vibrational populations~\cite{Gadzuk1991,SurSci1,SurSci2,Gao1997,Gao1997tunnel,Olsen2009}. This effect can be rationalized as a reduced dissociation barrier~\cite{Christopher2011,Christopher2012,Mukherjee2013,Nordlander2018,Nordlander2019,Nordlander2020} without considering the modified vibrational temperatures. However, a microscopic description of vibrational temperature presented here is more physical and aligns with experimental observations of elevated and mode-dependent vibrational temperatures~\cite{Linic2016_1,Linic2016_2,Dong2024,Frontiera2018,Kim2023,Kim2024}.
	
	The light intensity dependence of the dissociation rate suggests that the thermal contribution dominates only under low light irradiation, and is gradually prevailed by the nonthermal contribution at increasing light intensities. Such an unambiguous distinction between the thermal and nonthermal effects is crucial in understanding the mechanism involved, which hasn't been achieved in previous models~\cite{Olsen2009,Christopher2012}. It benefits from a unified treatment of the thermal and nonthermal hot carrier distributions. Moreover, the nonthermal hot electrons are found to be more favorable to trigger molecular dissociation with the corresponding rate being several orders of magnitude higher than the hot holes. This is consistent with the vibrational transition rates shown in Fig.~\ref{fig:fig3}, which originates from the better energy alignments between hot electrons and molecular resonance. We note that recent experiments proposed an alternative hot hole mechanism for single O$_2$ dissociation in a tunneling configuration,~\cite{Kim2020,Kim2023_2} which is interesting and deserves future investigation. Therefore, Fig.~\crefformat{figure}{#2#1{(b)}#3}\cref{fig:fig4} suggests that at relatively high light intensities, nonthermal hot carriers can greatly promote and even dominate the plasmonic photocatalysis, and the reaction mechanism evolves from single to multiple electronic transitions at the onset of significant vibrational heating.
	
\begin{figure*}
		\centering
		\includegraphics[width=3.3 in, keepaspectratio]{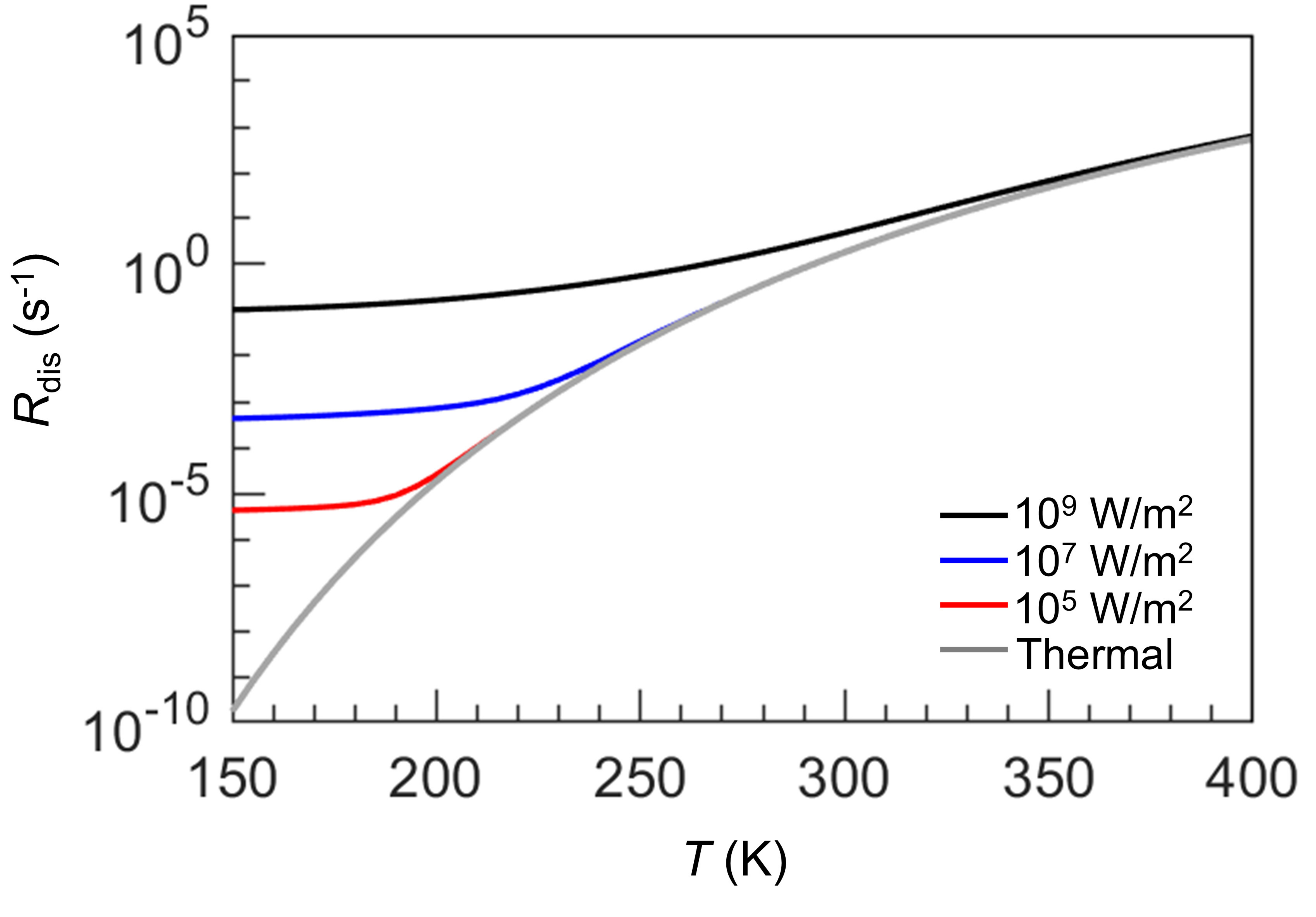}
		\caption{\label{fig:fig5} Temperature dependence of the total dissociation rate at different light intensities under resonant excitation. The nonthermal contributions are dominant in the low temperature regime as compared to the thermal reaction limit (gray solid line). All calculations are taken at $n_\mathrm{d}=6$, and $\tau=100\;\mathrm{fs}$.}
	\end{figure*}

	Figure~\ref{fig:fig5} shows the dissociation rate as a function of temperature in the vibrational heating mechanism. The thermal rate (Thermal) corresponds to a background dissociation rate under the dark condition. It essentially follows the Arrhenius law. Plasmonic hot carriers under light irradiation opens nonthermal channel of dissociations that increases with light intensity. Such a nonthermal enhancement is more dramatic in the low temperature regime, where thermal carrier excitation are suppressed and the dynamic activation to the dissociation level is thus only accessible through nonthermal hot carriers. As the temperature increases, thermal excitation of higher vibrational states becomes more efficient. It effectively diminishes the nonthermal counterpart. The results here are in line with early data in the DIET mechanism~\cite{Wu2022}, and expose additional light intensity dependence in the vibrational heating mechanism. Figure~\ref{fig:fig5} indicates that plasmonic hot carriers can promote photodissociation mostly in low temperature and high-light intensity regime, where vibrational heating induced by the nonthermal hot carriers becomes efficient and dominant.
	
	While the model captures essential feature of nonthermal hot carriers in the rate of plasmonic photodissociation and its transition from linear to superlinear light intensity dependence, the calculated threshold ($\sim10^8\;\mathrm{W/m^2}$) of the transition is considerably higher than the light intensity measured in the experiment ($\sim10^3\;\mathrm{W/m^2}$) \cite{Christopher2012}. This difference between theory and experiment could be attributed to the extra local field enhancement present at the plasmonic gaps in experiment. Typically, the experiment was performed on ensemble of nanocubes, where the photodissociation mostly take place at hot spots in narrow plasmonic gaps~\cite{Christopher2012}. The local field therein is substantially higher than the surface field on a single spherical nanoparticle as treated in our calculations~\cite{Nordlander2011}. For instance, the field enhancement at the hot spot of the nanoparticle dimer can be more than 3 orders of magnitude higher than that near an individual nanoparticle~\cite{Christopher2012}. This local field argument is also consistent with the experimental observation that single nanoparticle did not yield photodissociation at the same experimental conditions~\cite{Christopher2012}. In addition, dynamics beyond one-dimensional harmonic oscillator, and possible inhomogeneous density distribution of hot carriers~\cite{Erhart2020}, which are neglected in our model but may present in larger nanoparticles as used in experiment~\cite{Maier2017,Cortes2024,Cortes2024_2}, may be the additional factors contributing to the discrepancy.
	
	\subsection{Which hot carriers are operative?}
\begin{figure*}
		\centering
		\includegraphics[width=3.3 in, keepaspectratio]{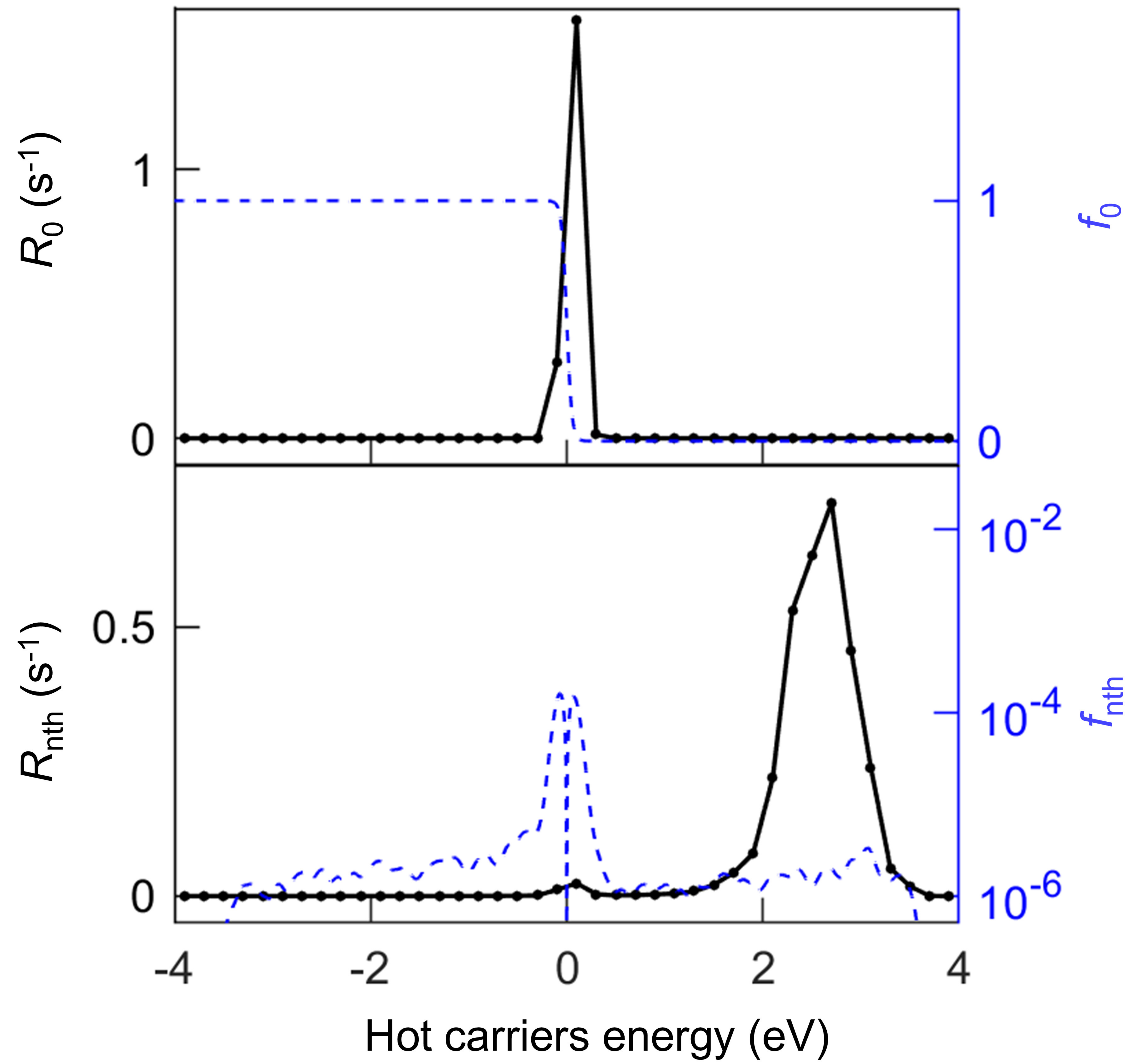}
		\caption{\label{fig:fig6} Thermal and nonthermal contributions to reaction rates obtained in different energy regions of the hot carriers spectrum. While the thermal contributions are dominated by hot carriers close to the Fermi level, nonthermal hot carriers with a broad energy range from near the molecular resonance to the plasmon resonance provide comparable contributions. Parameters used in the calculations are: $h\nu$=3.5~eV, $T=300\;\mathrm{K}$, $I_0=10^9\;\mathrm{W/m^2}$, and $\tau=100\;\mathrm{fs}$.}
	\end{figure*}

	This answer to this question is relevant to the mechanism of plasmonic photochemistry. Although the photodissociation of O$_2$ has been analyzed in a similar model~\cite{Olsen2009,Christopher2012}, it does not explicitly account for the hot carrier distributions and their energy dependent activation to a molecular resonance. Instead, it was assumed that a single-hot carrier at the plasmon resonance energy is exclusively involved in the scattering event. To gain insight into the which carriers are important for photocatalysis, we have segmented the dissociation rates in the hot carrier spectrum with an energy step of 0.2~eV. As shown in the upper panel of Fig.~\ref{fig:fig6}, in the thermal regime, only the carriers near the Fermi level provide the dominant contributions to the molecular dissociation. This results from the electron-hole pair distribution factor, $f(\varepsilon)[1-f(\varepsilon+\delta\varepsilon)]$ in Eq.~\eqref{eqn6}, which is peaked at the Fermi energy of the thermal electronic bath. Therefore, the low energy electron-hole pairs near the Fermi level are the most active carriers in activating the molecules through an off-resonance excitation mechanism. In contrast, the nonthermal distribution $f_\mathrm{nth}$ has appreciable distributions in the high energy regime. These energetic carriers better align with the 2$\pi^*$ resonance of the O$_2$ molecules, dominating the contributions to the molecular dissociation through a resonance electron-molecule excitation.
	
	A detailed look at the resonance dissociation peak in the lower panel of Fig.~\ref{fig:fig6} suggests that the hot carriers involved can not be attributed to a single electron excitation with energy equal to neither the plasmon energy (3.5~eV~\cite{Christy1972}) nor the molecular resonance (2.4~eV~\cite{Christopher2012}). Instead, it shows a broad energy distribution from 1.8-3.6~eV, demonstrating the combined effects of the molecular resonance and the nonthermal hot carrier distributions. In other words, such a broad energy profile results from the interplay between molecular resonance and plasmonic hot carrier distributions. It suggests that a proper treatment of the energy-dependent coupling with the hot electron bath is important to understand the mechanism and is an essential step beyond a single-electron scattering picture~\cite{Christopher2012,Olsen2009}. The unified description of the thermal and nonthermal hot carriers in our model provides a concise theoretical framework and quantitative assessment of their relative contributions in the coupling with the full bath of hot carriers. It offers deeper insight into the mechanism of plasmonic photocatalysis within a well-defined model.

	\section{Conclusion}\label{Con}
	Based on the Anderson-Newns model of resonance electron-vibration coupling and an explicit derivation of nonthermal hot carrier distribution, we have presented a unified description of vibrational heating and dissociation induced by plasmonic hot carriers, with the focus on the mechanism of plasmon induced photocatalysis. As an application of the model, we revisited the photodissociation of O$_2$ adsorbed on AgNP and systematically examined the dissociation rate as functions of temperature and light intensity. Treating the nonthermal and thermal distribution on equal footing has allowed us to gain quantitative assessment of the contributions of nonthermal carriers in comparison with that of thermal carriers. The results demonstrate that a small fraction of population by energetic nonthermal carriers is sufficient to greatly enhance the vibrational excitation, thus promoting the catalytic dissociation, especially in the low temperature and/or the high light intensity regime. Furthermore, single electronic scattering dominates the dissociation under weak excitation, vibrational heating induced by multiple electronic scatterings outweighs as the light intensity increases, giving a linear to superlinear dependence of dissociation rate as found in experiment. Our model captures the essential physics underlying the mechanism of plasmonic catalysis, and provides a unified theoretical framework to describe hot carrier photochemistry in both the thermal and nonthermal regimes. 	
	\begin{acknowledgement}
		This work was supported by the National Natural Science Foundation of China (11934003, 12393831, U2230402).
	\end{acknowledgement}
	
		
		
		
	
	\bibliography{ref}
	
\end{document}